\documentclass[11pt,superscriptaddress]{article}
\pdfoutput=1

\usepackage{cite}	
\usepackage{enumerate}
\usepackage{empheq}
\usepackage{booktabs}
\usepackage[english]{babel}
\usepackage{amsmath,amssymb,amsbsy,amstext, amsthm, simplewick}
\usepackage{graphicx}
\usepackage{amsfonts}
\usepackage{color}
\usepackage{framed}
\usepackage[top=3cm, bottom=3cm, left=3cm, right=3cm]{geometry}
\usepackage{verbatim}

\usepackage{authblk}

\usepackage{colortbl}
\definecolor{summersky}{cmyk}{0.71,0.33,0,0.5}
\definecolor{flamingo}{cmyk}{0,0.51,0.71,0.5}
\definecolor{rp}{cmyk}{0.2, 1, 0.6, 0}
\definecolor{pacificblue}{cmyk}{0.95,0.3,0, 0.5}
\definecolor{gray60}{cmyk}{0.4,0.4,0,0.8}

\newcommand{\ex}[1]{\langle #1 \rangle}

\newcommand{\be}{\begin{eqnarray} }
\newcommand{\ee}{\end{eqnarray} }
\newcommand{\bs}{\begin{split} }
\newcommand{\es}{\end{split} }

\newcommand{\Mpl}{M_{\mathrm Pl}}

\def\be{\begin{eqnarray}}
\def\ee{\end{eqnarray}}

\def\[{\left [}
\def\]{\right ]}
\def\({\left (}
\def\){\right )}

\def\r2{\sqrt{2}}

\def\O{{\mathcal O}}
\def\Mpl{M_{\mathrm Pl}}
\def\lmin{\lambda_{\mathrm{min}}}
\def\fvb{ \boldsymbol{\bar{ \phi} } }
\def\fv{ \boldsymbol{ \phi} }

\graphicspath{{Figures/}}

\begin{document}

\begin{titlepage}

\bigskip\

\vspace{.5cm}
\begin{center}

{\fontsize{25}{28}\selectfont  \sffamily \bfseries  Inflation on a Slippery Slope}


\end{center}

\vspace{1cm}

\begin{center}
{\fontsize{13}{30}\selectfont Ben Freivogel$^1$, Roberto Gobbetti$^{1, 2}$ , Enrico Pajer$^{2}$, and I-Sheng Yang$^{3, 4}$}
\end{center}


\begin{center}

\vskip 8pt
\textsl{$^1$ GRAPPA and ITFA, Institute of Physics, Universiteit van Amsterdam, \\
Science Park 904, 1090 GL Amsterdam, Netherlands}
\vskip 7pt

\textsl{$^2$ Institute for Theoretical Physics and\\
Center for Extreme Matter and Emergent Phenomena,\\ Utrecht University, Princetonplein 5, 3584 CC Utrecht, The Netherlands}
\vskip 7pt

\textsl{$^3$ Perimeter Institute of Theoretical Physics, \\
31 Caroline Street North, Waterloo, ON N2L 2Y5, Canada,}
\vskip 7pt
 
\textsl{$^4$ Canadian Center of Theoretical Astrophysics, \\
60 St George St, Toronto, ON M5S 3H8, Canada}

\end{center}

\vspace{2cm}
\hrule \vspace{0.3cm}
\noindent {\sffamily \bfseries Abstract} \\

\noindent We study inflation in a random multifield potential, using techniques developed by Marsh {\it et al}.  The potential is a function of a large number of fields, and we choose parameters so that inflation only occurs in regions where the potential is accidentally flat. Using an improved estimate for the dynamics of eigenvalue repulsion, we are able to describe the steepening of the potential as inflation progresses. We provide suggestive arguments, but not a proof, that the resulting scalar power spectrum generically disagrees with observations. We also point out two problematic aspects of the model: there is no well-defined probability distribution for the gradient of the potential, and the evolution of the potential over small distances in field space is unphysical.

\vskip 10pt
\hrule

\vspace{0.6cm}
 \end{titlepage}

\tableofcontents

\section{Introduction}

The inflationary paradigm provides a successful model for the origin of inhomogeneities in our universe \cite{Guth:1980zm,Linde:1981mu,Albrecht:1982wi}. The picture painted by the latest data is consistent with single field models of inflation, and all we know can be described by a nearly scale invariant two-point function. In other words, our universe looks very simple \cite{Ade:2015lrj}.

On the other hand, the UV theories at our disposal are complex. Following the renormalization group flow in reverse, one expects to find more and more degrees of freedom at high energies. String theory predicts the existence of many fields and a complicated landscape of solutions which no one has been able to fully characterize \cite{BP}.

The apparent clash between observation and high energy theories leaves us with two options: either we should completely revisit the way we model UV physics, or we need to understand how lots of degrees of freedom conspire to produce a simple universe. In this paper, we will pursue the second possibility. Building on previous work, we present a way to model a multifield landscape and study whether it leads to an inflation era consistent with observations. 

The natural way to describe the behavior of many degrees of freedom is to coarse-grain them and use statistics. This approach has a long history \cite{Wigner:rm,Dyson:bm}. For a multifield potential, the usual procedure is to expand a potential at second order and extract a random matrix that represents the couplings between the fields \cite{Easther:2005zr}. Even though we do not know the actual UV theory in detail, the statistics only depends on the choice of a few parameters. Such a statistical description of the landscape provides a framework to tackle problems like computing the probability of having stable minima \cite{Denef:2004ze,MarMcA11,Chen:2011ac,Bachlechner:2012at,GreKag13,Bachlechner:2014rqa}, or the probability of inflation occurring \cite{McARen12,Yan12a,Marsh:2013qca,Pedro:2013nda}. In this work we take a step further along the second line by addressing the question: do the statistical predictions agree with cosmological observations?

We will follow the prescription delineated by Marsh, McAllister, Pajer and Wrase in \cite{Marsh:2013qca} (henceforth MMPW) to model a random landscape. Unless the potential is slow-roll flat almost everywhere, Inflation typically occurs in regions of the potential where the slope is ``accidentally" flat. A flat potential requires that no field has a very negative mass squared --- in other words, inflation occurs in regions of the potential in which none of the eigenvalues of the second derivative matrix is too negative. As inflation proceeds, the inflaton moves through field space and the eigenvalues evolve. Assuming that the entries of the second derivative matrix are statistically independent leads to the phenomenon of {\it eigenvalue repulsion}: the lowest eigenvalue is repelled by the other eigenvalues, and quickly becomes more negative. 

We give an improved analysis of how the lowest eigenvalue evolves over short distances in field space, finding a field-dependent mass term of the form
\be
m^2(\phi) = m_0^2 - c \phi^{2/3}
\ee
This effect causes a steepening of the potential, leading to inflation on a ``slippery slope."  

 Calculating the perturbations correctly is subject to a number of uncertainties in how the additional degrees of freedom evolve through inflation and reheating. Here we will do the only simple calculation available, and compute the perturbations within the effective single field model which was shown by MMPW to be a good description of the background evolution.
 
We provide strongly suggestive arguments, but not a proof, that the predictions of such models disagree with data. We analyze the case where inflation occurs near an exact critical point, where the gradient is zero. After conditioning on the length of inflation and the amplitude of density perturbations, we find that the spectral index is too red to agree with observation. This problem can be alleviated by inflating from a point with small gradient which is \textit{not} near any critical point.  We believe that this alternative is finely tuned in the choice of the gradient. Unfortunately, as we discuss below, the statistics of the gradient is ill-defined in the MMPW model and hence we cannot quantify the degree of fine tuning.

Our results also rule out a wider class of models than the particular implementation of MMPW: any effective single-field model of hilltop inflation with a field-dependent mass term,
\be \label{gen_potential}
	V(\phi) = V_0  - \left({1 \over 2} m_0^2 + c \phi^p \right) \phi^2
\ee
with positive $c$ and $0 < p < 3$ is in conflict with data as long as inflation ends in the regime where the $\phi^p$ term dominates over the bare mass term $m_0^2$. We will show that we can expect field evolution in random landscape models near a critical point to be described by equation \eqref{gen_potential} and to end well into the $\phi^p$ regime, at least in the approximation of gradient flow evolution.

Treating the multifield landscape as a single field model is clearly an approximation as we are disregarding any possible multifield effects apart from eigenvalue repulsion. Nevertheless, the full multifield behavior would not invalidate our arguments as long as the turning frequency is not too high compared to the horizontal scale of the potential $\Lambda_h$. It is interesting to ask whether the additional degrees of freedom would significantly modify our results. For an interesting insight on this issue, see \cite{Dias:2016slx} and upcoming work from the same authors.

In the process of performing this analysis, we have run into two problems with the MMPW model. First, the evolution of the Hessian in field space is unphysical for small field displacements. The problem is that the model describes the evolution of the Hessian by diffusion, meaning that for small distances in field space $\delta V_{ab} \sim \sqrt{\delta \phi}$, where $V_{ab}$ is an entry in the Hessian matrix. However, as long as third derivatives of the potential are finite, the evolution over small distances must be linear, $\delta V_{ab} \sim \delta \phi$.  Therefore, the MMPW model predicts too large of a change in the Hessian over small distances in field space. This issue is important for the predictions of the model, since much of inflation occurs over short distances in field space. This issue has been raised previously by Battefeld and Modi \cite{Battefeld14}.

The second problem is that the predictions depend sensitively on the probability distribution for the potential energy and its gradient. While MMPW showed that there is a probability distribution for the Hessian that is invariant under the evolution in field space, one can show that there is no probability distribution for the gradient that is invariant under the MMPW evolution equations. This has been shown previously in unpublished work by Marsh and Pajer \cite{MarshPajer}, and we give a sketch of the argument in section \ref{ssec-gradient}.

Both of these issues are discussed in the recent work of Battefeld and Wang \cite{Battefeld16}. A related discussion of the issue of graceful exit in the MMPW model recently appeared in \cite{Pedro}.

 This paper is structured as follows: In section \ref{review} we review the stochastic model proposed in MMPW \cite{Marsh:2013qca} and the random matrix theory tools needed in this work. In section \ref{rando}, we derive a new and more accurate description of the evolution of the smallest eigenvalue of the mass matrix for small field displacements. In section \ref{tilt_comp}, we calculate the observational predictions of the model within the single-field approximation. 
 We explain the problems with the MMPW model in section \ref{issues}. We conclude in section \ref{discussion} with a discussion of the validity and implications of our results.


\section{A random potential from Dyson Brownian Motion} \label{review}

In this section, we critically review the model in MMPW \cite{Marsh:2013qca}. Besides introducing our notation and reviewing the necessary mathematical background, we clarify two important points: the total number of physical parameters and the freedom in the choice of the dynamical evolution of the Hessian.

The approach proposed in MMPW to generate a random landscape is perhaps the most pragmatic. Instead of modeling the landscape in the whole large multifield volume, only small regions around a given trajectory are constructed. We review now the construction, dividing it into two logical steps. First, we discuss the \textit{local} statistical properties of the potential, i.e.~around an arbitrary point. Second, we show how to dynamically evolve the random potential along an arbitrary trajectory.
As we will discuss, the advantage of the model is that it is tractable, and it does not make \textit{ad hoc} assumptions about the potential. We will discuss later some ways in which this approach might be improved. 


\subsection{The local description}

At any given point $ \fvb \in \mathbb{R}^{N_f}$, with $N_f$ to total number of active scalar fields, one Taylor expands the potential to second order as
\be \label{expansion}
	V = \Lambda_v^4\,\sqrt{N_f}\[v_0 + v_a\frac{\phi^a}{\Lambda_h} +\frac{1}{2} v_{ab}\frac{\phi^a\phi^b}{\Lambda_h^2}+\dots\] \, ,
\ee
where $\Lambda_v$, $\Lambda_h$ and $N_f$ are ``global'' parameters of this landscape and do not depend on the chosen point $ \fvb $. They specify the vertical and horizontal (mass) scales, and the total number of fields respectively. $v_0(\fvb)$, $v_a(\fvb)$ and $v_{ab}(\fvb)$ are local random variables, which determine the randomness of $ V $. As it will become clear later, the peculiar factor $ \sqrt{N_f} $ was chosen for convenience\footnote{We will eventually take $v_{ab}\sim\mathcal{O}(1/\sqrt{N_f})$ to ensure its eigenvalues are typically $\mathcal{O}(1)$.}. We make three further assumptions: (i) $ V(\fv) $ is statistically homogeneous and isotropic; (ii) $ v_{0} $, $ v_{a} $ and $ v_{ab} $ are independent random variables for each $ a $ and $ b $; (iii) all random variables are Gaussian with zero mean. In the limit of large $ N_f $, several properties of the potential arise that do not depend on the probability distribution of the individual random variables\footnote{This is true provided the probability distribution satisfies some regularity criteria, such as exponential tails that ensure the finiteness of all cumulants.}, so assumption (iii) is not as restrictive as one might suspect. This universality of the limit of a large number of fields, $ N_f\gg 1 $, is what makes this approach particularly attractive. 

As a consequence, the random, symmetric, square matrix $v_{ab}$ must be a realization of the Gaussian Orthogonal Ensemble (GOE). Namely, the resulting $ N_f(N_f+1)/2 $ entries are statistically independent Gaussians that obey the probability distribution
\be
P_{\mathrm{GOE}}(v_{ab})=\mathcal{C}\exp\left[ - \frac{\mathrm{Tr}\left(\boldsymbol{v}^{2} \right)}{2\sigma^{2}}\right]=\mathcal{C}\exp\left[ - \frac{\sum_{a,b} \left( v_{ab} \right)^{2}}{2\sigma^{2}}\right]\,,
\ee
with $ \mathcal{C} $ a normalization factor, and where we have assumed zero average and some (dimensionless) variance $ \sigma $. The relevant expectation values are then
\be\label{fact2}
\ex{v_{ab}}=0\,,\quad \langle  v_{ab}^2 \rangle = \sigma^{2} \frac{(1 + \delta_{ab})}{2}\,.
\ee

A well-known result in Random Matrix theory is that the eigenvalues $ \lambda_{i} $ of $ v_{ab} $ are distributed according to the Wigner semicircle law, i.e.~the probability density for any one eigenvalue $\lambda$ is 
\be
	\rho(\lambda) = \frac{1}{\pi N_f \sigma^2}\sqrt{2N_f\sigma^2-\lambda^2} \, .
\ee

Altogether, we have now naively six parameters
\be
\left\{\Lambda_{v},v_{0},v_{a},v_{ab},\Lambda_{h},N_f\right\}\,,
\ee
where $v_{0}  $, $ v_{a} $ and $ v_{ab} $ stand here for the variance of their respective Gaussian distribution\footnote{Notice that, because of statistical isotropy, all $ v_{a} $ and all $ v_{ab} $ must have the same variance for every $ a,b $ (up to the factor of 2 in \eqref{fact2}), so they count as two parameters rather than $ N_f $ and $ N_f(N_f+1)/2 $, respectively.} (which we called $ \sigma $ for $ v_{ab} $ in \eqref{fact2}). This parameterization is clearly redundant. The potential in \eqref{expansion}, thought of as a function $ V(\phi):\mathbb{R}^{N_f}\rightarrow \mathbb{R} $, is invariant under two independent scalings, with charges\footnote{The numbers refer to the exponent of the scaling $ \gamma $ by which the $ i $-th parameter is changed. For example, for the first scaling $ \Lambda_{v}\rightarrow \gamma^{1} \Lambda_{v} $, $ v_{0,a,bc}\rightarrow \gamma^{-4} v_{0,a,bc} $ and so on.}
\be
\{1,-4,-4,-4,0,0\}\quad \mathrm{and} \quad \{0,0,1,2,1,0\} \,.\label{scalings}
\ee 
Notice that $ N_f $ does not participate in the scalings because a change in $ N_f $ changes the dimensionality of the ambient space, and clearly can never leave $ V(\phi) $ invariant. Also, the normalization of $ \phi $ is fixed by the requirement of a canonical kinetic term. One could use the two scalings to fix two of the six parameters. Before doing this though, we have to construct the random potential away from a single point.


\subsection{The dynamical description}

To describe the dynamics of inflation, we need to define the potential along some trajectory in field space. The proposal of MMPW is to dynamically evolve the local approximation of the previous subsection, as we move through field space. Under an infinitesimal displacement $\delta \fv  $ from some given point $ \fvb $, the first two Taylor coefficients evolve in the usual way, namely
\begin{eqnarray}
\delta v_0\equiv v_{0}(\fvb+\delta \fv)-v_{0}(\fvb)= v_a \frac{\delta \phi^a}{\Lambda_{h}} \,,\\
\delta v_a \equiv v_{a}(\fvb+\delta \fv)-v_{a}(\fvb) = v_{ab} \frac{\delta \phi^b}{\Lambda_{h}}\,.
\end{eqnarray}
Notice that this evolution is invariant under both scalings \eqref{scalings}. To avoid an infinite set of equations for the infinite set of Taylor coefficients, MMPW assumed that the matrix of second derivatives, a.k.a.~the Hessian, evolves according to a very specific stochastic process: Dyson Brownian Motion \cite{Dyson:bm}. Namely, for an infinitesimal displacement  $ ||\delta  \fv || \rightarrow 0 $, $ v_{ab}(\fvb+\delta \fv)=v_{ab}(\fvb)+\delta v_{ab}(\fvb) $ and $ \delta v_{ab}$ is given by the following simple stochastic process
\be\label{evolution}
	\begin{split}
	\langle \delta v_{ab}(\fvb) \rangle &= - v_{ab}(\fvb) \,\frac{||\delta  \fv ||}{\Lambda}\,, \\
	\langle \delta v_{ab}(\fvb)^2 \rangle &=  \sigma^{2} \frac{(1 + \delta_{ab})}{2} \,\frac{||\delta  \fv ||}{\Lambda}\,,
	\end{split}
\ee
where $ \Lambda $ is a new mass scale and $ \sigma $ is the variance of $ v_{ab} $ as in \eqref{fact2}. Intuitively, the first equation is a restoring force that tends to re-center the matrix entries around 0, while the second term describes diffusion of the matrix elements as we move through field space. 

Besides simplicity, the choice of this particular stochastic process is motivated by the fact that \textit{the GOE is the asymptotic, ``stationary'' solution of \eqref{evolution} for the probability distribution of $ v_{ab} $}. This can be seen by deriving the Fokker-Planck-Smoluchowski equation for the system and studying the ``time''-independent solution, where $ \phi $ plays the role of time. In other words, assuming Dyson Brownian motion and moving away from some $ \fvb $ along some trajectory, one finds that $ v_{ab}(\fv) $ becomes a realization of the GOE for $ ||\fv-\fvb|| \gg \Lambda $. This happens independently of the initial probability distribution for $ v_{ab}(\fvb) $. Because of this, $ \Lambda $ is interpreted as the correlation length of the random potential. Points at distance larger than $ \Lambda $ have uncorrelated Hessians.

Two important shortcomings of this procedure are that the Hessian evolution is unphysical for very small displacements and the probability distribution of $ v_{0} $ and $ v_{a} $ do not have an asymptotic ``static'' solution. We defer a discussion of these issues to section \ref{issues}. 

The stochastic process in \eqref{evolution} is invariant under the two scalings in \eqref{scalings} as long as we postulate that $ \Lambda $ has weight zero in both cases. It is convenient to use the second scaling to fix 
\be
	\Lambda_{h}\overset{!}{=}\Lambda \,.
\ee 
This is because, from the point of view of a Wilsonian effective action, one naturally thinks of $ \Lambda_{h} $ as the cutoff of the effective theory around some $ \fvb $, or in other words, the radius of convergence of the Taylor expansion. But from the discussion above, we established that $ \Lambda $ is the distance beyond which the Taylor expansion is uncorrelated with the initial one. Without loss of generality, we can therefore impose the ``natural'' choice $ \Lambda_{h}=\Lambda $. Finally, we fix the second and last scaling symmetry by choosing
\be
	\sigma^{2}\overset{!}{=}\frac{2}{N_f}\,.
\ee
This ensures that the eigenvalues $ \lambda_{a} $ of $ v_{ab} $ are distributed in the interval $ \left\{ -2,2\right\} $. Summarizing, the full model has 5 parameters: $ \Lambda_{v} $, $ v_{0} $, $ v_{a} $, $\Lambda_{h}  $ and $ N_f $.

 
\subsection{Regimes of inflation}

We are interested in studying slow-roll inflation in the MMPW model. Generally speaking, the first condition for slow-roll inflation to take place at all is a slowly changing expansion rate, so
\begin{eqnarray}
\epsilon \equiv -\frac{\dot{H}}{H^2} \approx \frac{3\dot{\phi}_a\dot{\phi^a}}{2V}
\ll 1~.
\end{eqnarray}
In the single field case, the equation of motion for the field can only sustain such condition if the $\ddot{\phi}$ terms is negligible, namely $3H\dot\phi = -V'$. In this paper, we will adopt its multifield generalization: $3H\dot{\phi}_a = -V_a$, where the subscript on $V$ denotes its partial derivative with respect to a scalar field component. This condition means that the slow-roll trajectory follows the gradient flow of the potential. Although multifield inflation can happen without such a condition \cite{Achucarro:2010jv,Yang:2012bs,Ahlqvist:2013wla}, there is not yet a strong reason why gradient-flow inflation is statistically disfavored. Besides, the observational consequences of non-gradient-flow inflation are much harder to study in general (see e.g.~\cite{Achucarro:2010da,Achucarro:2012sm,Avgoustidis:2012yc}), thus we will not include them in this paper. 

Given the gradient-flow condition, we can write the first slow-roll parameter entirely as a function of the potential,
\be
\epsilon_{V}&=&\frac{\Mpl^{2}}{2}\frac{||V_{i}||^{2}}{V^{2}}=\frac{1}{2}\frac{\Mpl^{2}}{\Lambda_{h}^{2}}\frac{||\left(  v_{a}+v_{ab}\phi^{b}/\Lambda_{h}\right)||^{2}}{\left(v_{0}+   v_{a}\phi^{a}/\Lambda_{h}+v_{ab}\phi^{a}\phi^{b}/\Lambda_{h}^{2}\right)^{2}}\ll1\,,
\ee
so that one can directly study its statistical properties. 

The second slow-roll condition is basically that the first condition lasts sufficiently long to produce a spectrum of scale invariant perturbations. The condition for a slowly changing $\epsilon_V$, under the gradient flow condition, is given by \cite{Yang:2012bs}
\be
\eta&\equiv&\frac{V_{i}}{||V_{l}||} \frac{\Mpl^{2}V_{ij}}{V} \frac{V_{j}}{||V_{l}||},\quad |\eta|\ll 1\,.
\ee
However, the gradient flow condition, which we will assume throughout this paper, is already a stronger requirement \cite{Yang:2012bs},
\be
\xi&\equiv&\sqrt{\frac{V_{i}}{||V_{l}||} \frac{\Mpl^{2}V_{ij}}{V}\frac{\Mpl^{2}V_{jk}}{V} \frac{V_{k}}{||V_{l}||}}\ll 1\,.
\ee

$ \xi $ and $ \eta $ become more transparent in a basis in which the matrix $ \Mpl^{2}V_{ij}/V $ is diagonal with eigenvalues $ \tilde\lambda_{i} $ (where a tilde distinguishes it from the eigenvalues $ \lambda_{i} $ of $ v_{ab} $). In that basis, following \cite{Yang:2012bs}, define the normalized gradient vector $ V_{i}/||V_{l}||=n_{i} $, thus $ \sum_{i}n_{i}^{2}=1 $. We can write down
\be
\eta=\sum_{i}n_{i}^{2}\tilde\lambda_{i}\quad\mathrm{and}\quad \xi=\sqrt{\sum_{i}n_{i}^{2}\tilde\lambda_{i}^{2}}\,.
\ee
One recognizes $ \eta $ as the weighted arithmetic mean and $ \xi $ as the weighted root mean square of the (rescaled) Hessian eigenvalues. By the generalized weighted mean inequality $ \xi>|\eta|>0 $, so indeed gradient flow implies slow-roll inflation but not viceversa.

Given all the assumptions we stated in this section, we find three regimes in which slow-roll inflation can happen:
\begin{itemize}
\item $\Lambda_h\gg M_p$, i.e.~the potential is very flat for typical values of $ v_{0}\sim ||v_{a}||\sim \O(1) $; 
\item $ v_0\gg1$, i.e.~the slow-roll parameters are suppressed by the high scale of inflation;
\item both the norm of the gradient $ v_{a} $ and the smallest eigenvalue $ \lmin $ of $ v_{ab} $ are much smaller than one.
\end{itemize} 
In the first two regimes, inflation can always parametrically happen and we do not discuss them further in the remaining of the paper. We instead concentrate on the third possibility. Moreover, we assume that inflation happens within one correlation length, namely with a field excursion smaller than or comparable with $\Lambda_h$. This is not strictly necessary, but it is probabilistically more difficult to sustain inflation within several patches, and the problems that affect single patch inflation affect a longer excursion as well.

 
\section{Inflation in a random potential}
\label{rando}

In this section, we review the phenomenon of eigenvalue repulsion, introduce the effective single-field description and derive a new approximation for the evolution of the smallest eigenvalue for small field displacements. The comparison with numerical results confirms the accuracy of our approximation.

\subsection{Eigenvalue distribution} \label{sec:relaxation}

To study inflation near a critical point, we are interested in the masses of the fields near that point, which are related to the eigenvalues of the the Hessian $\lambda_i$ through
\be
	m_i^2 = \frac{\Lambda_v^4}{\Lambda_h^2}\sqrt{N_f}\lambda_i \, .
\ee

\begin{figure}[t]
	\centering
	\includegraphics[width=.8\textwidth]{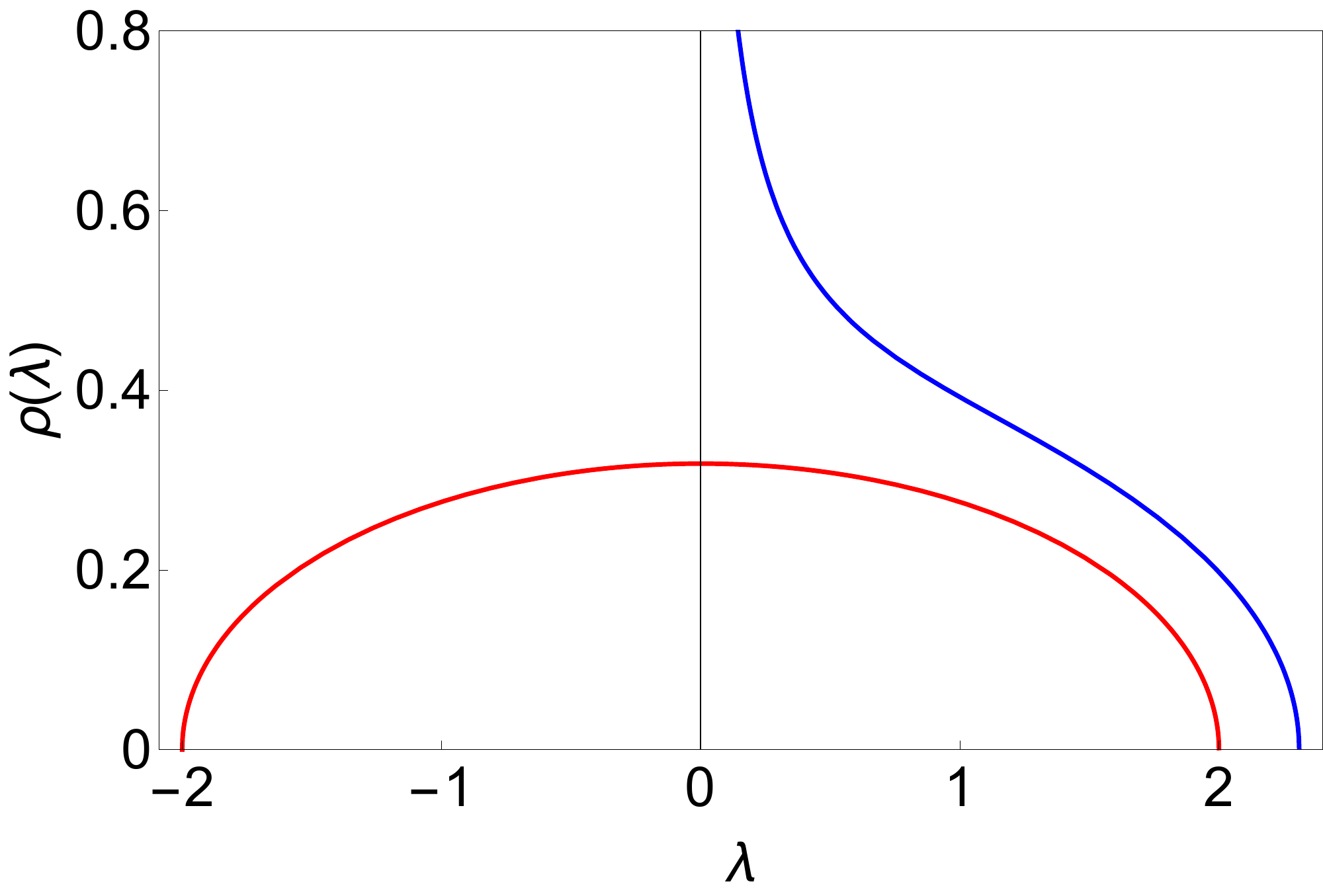}
	\caption{\label{fluctuation}We plot here the Wigner semicircle distribution for the eigenvalues of a random matrix (red) compared to the fluctuated distribution needed to ignite inflation at a critical point (blue).}
\end{figure}

The Wigner distribution implies that, for a typical realization, about half of the directions in field space are tachyonic. To realize slow-roll inflation in this landscape, we must postulate that our universe is the result of a fluctuation away from the Wigner distribution, so that all fields have masses above some small negative value. The probability of such a fluctuation is exponentially suppressed by the number of fields,
\be
	P(\lambda_{min}\geq\xi)\sim e^{-N_f^2\xi}\,,
\ee
for $|\xi|\ll1$. In figure \ref{fluctuation} we show the full probability distribution requiring only positive eigenvalues (blue line) versus the Wigner semicircle (red line).

As the name suggests, the fluctuated probability distribution is not an equilibrium distribution. To understand what we mean by this, it is useful to interpret the eigenvalues as a monodimensional gas of particles subject to the potential
\be \label{wigner_potential}
	W = \frac{1}{\sigma^2}\sum_{i=1}^{N_f}\lambda_i^2 - \sum_{i\neq j}\ln\left|\lambda_i-\lambda_j\right| \, .
\ee
It can be shown \cite{mehta} that the equilibrium of this system leads precisely to the Wigner distribution. This analogy is useful to understand the situation out of equilibrium: the eigenvalues experience the force due to the potential in equation \eqref{wigner_potential} and relax to the Wigner distribution in a given ``time scale''\footnote{Dyson showed \cite{Dyson:bm} that this is not quite enough to understand the dynamics of the system. One needs to introduce a stochastic term so that the system effectively behaves as a random walk with a drift term. Nevertheless, when the distribution is very skewed as in our case and the average distance between two eigenvalues is very small, the argument delineated here is valid.}.

In the MMPW model, the time scale is the field displacement relative to the horizontal scale, $ \phi/\Lambda_h$. This implies that, as the system evolves, the potential receives corrections until the eigenvalues are distributed according to the Wigner distribution. This happens for a displacement $\Delta\phi\sim\Lambda_h$, which represents the correlations scale in field space.

\subsection{Effective single field description}

Let us consider a fluctuation where most of the eigenvalues are positive and only one (or a few) is negative and close to zero. In other words, calling the most negative eigenvalue $\lambda_1<0$, we require that 
\be
	|\lambda_1| \ll 1\, .
\ee

As shown \cite{Marsh:2013qca}, inflation can then be modeled by an effective single field potential 
\be \label{hilltop_potential}
V = V_0 +\gamma \phi  - \left[ \frac{m_{0}^{2}}{2}+\tilde{m}(\phi)^{2} \right]\phi^2  \, .
\ee
The mass term has acquired an additional $ \phi $-dependent to account for the effect of eigenvalue repulsion, which effectively steepens the potential as the inflaton evolves. The rest of this section is devoted to a discussion of what function $ \tilde{m}(\phi) $ best describes the actual multifield dynamics.


Defining
\be \label{rescaled}
	s \equiv {|| \fv || \over \Lambda_h} \, ,
\ee
MMPW estimated $ \tilde{m}(\phi) $ from an analytic lower bound, which leads to 
\be
\tilde{m}^2 \gtrsim 2(1-e^{-s})^{1/2}\,.
\ee
Then argued that a better fit to numerical results is obtained by the phenomenological expression
\be
\tilde{m}^2 \sim 2(1-e^{-s/2})^{1/2} \sim \sqrt{2}s^{1/2}\quad\text{(MMPW)}\,.
\ee
In the next subsection we derive analytically the scaling of $\tilde{m}$ in terms of $s$ and compare it with both simulations and the MMPW result.


\subsection{Modeling small field displacements} \label{sec:small_displacement}

We can derive a better estimate of how the eigenvalues evolve for small field displacement. For small $s$ and using perturbation theory, we show that $\lambda \sim - s^{2/3}$. This is a slower evolution of the mass than what found in MMPW, namely $\lambda \sim - s^{1/2}$. Comparison with numerical calculations show that our result is very accurate for small $ s $.

For small field displacement, the evolution in the MMPW model, defined by (\ref{evolution}), is dominated by diffusion,
\begin{equation}
\langle (\delta v_{ab})^2 \rangle = {1 \over N_f} (1 + \delta_{ab} ) \delta s\,,\ \ \ \ \ \ \langle \delta v_{ab} \rangle \sim s \approx 0 \,.
\end{equation}
This can be extended from infinitesimal $\delta s$ to finite but small $s$. The total change $\Delta v_{ab}$ is a sum of an infinite number of infinitesimal changes. Therefore, the total change is a gaussian random variable, with variance equal to the sum of the variances,
\begin{equation}
\langle (\Delta v_{ab})^2 \rangle = {1 \over N_f} (1 + \delta_{ab} )  s \ \ \ \ \ \ \ \  \ \langle \Delta v_{ab} \rangle \approx 0 \,,
\end{equation}
as expected for diffusion. The higher moments are determined, because $\Delta v_{ab}$ is Gaussian.

We would like now to estimate how the lowest eigenvalue depends on $s$, for small $s$, since, as we shall see later, all of inflation occurs in fact in the small $s$ regime. The full matrix of second derivatives is
\be
v_{ab} (s) = v_{ab} (0) + \Delta v_{ab}(s)\,.
\ee
Naively, we can expand using perturbation theory, as in quantum mechanics. Labeling the lowest eigenvalue $\lambda_1$, we have 
\be
\lambda_1 = \lambda_1(0) + \Delta v_{11} + \sum_{i = 2}^{N_f} {(\Delta v_{1i})^2 \over {\lambda_1 - \lambda i}} + ...\,.
\ee
However, recall that we are interested in the evolution near a critical point where all of the eigenvalues are constrained to be ``not too negative''. The eigenvalues at such critical points tend to bunch up near the minimum allowed value due to level repulsion. As a result, the denominators $\lambda_1 - \lambda_i$ are small for small $i$. In fact, $\lambda_2 - \lambda_1 \sim N_f^{-2}$. 

This means that we  have to do degenerate perturbation theory in a sector of the matrix. We can determine this sector through the following logic. Taking an $n$ by $n$ submatrix of the full Hessian, we compare the eigenvalues of $\delta v$ to the level spacing of $v_{ab}(0)$. The entries of $\Delta v$ are of size $\sqrt{s/N_f}$. The eigenvalues of an $n$ by $n$ matrix are of order $\sqrt n$ times the size of the entries, so the eigenvalues of the $\delta v_{ab}$ submatrix are of order $\sqrt{s n/N_f}$. 

In order to find $n$, notice that the eigenvalues of the unperturbed matrix are given roughly by 
\be 
\lambda_i - \lambda_1 \sim {i^2 \over N_f^2}\,.
\ee
We need to apply degenerate perturbation theory to a submatrix of size $n$ such that the eigenvalues of the submatrix are greater than or equal to the eigenvalue differences of the original matrix. This yields the equation for $n$
\be
{n^2 \over N_f^2} \sim \sqrt{s n \over N_f}\,,
\ee
which is solved by $n \sim N_f s^{1/3}$.

We can now compute the new lowest eigenvalue by combining the lowest eigenvalue of the degenerate matrix with the nondegenerate corrections. The $n \times n$ matrix $\delta v_{ab}$ will itself obey a semicircle law, with lowest eigenvalue $\sqrt n$ times the size of the entries, giving $- \sqrt{s n/N_f}$. Plugging in the value for $n$ given above gives $\lambda_1 \sim - s^{2/3}$. Calculating the additional contribution from the nondegenerate terms gives
\be
	\langle \Delta \lambda_1 \rangle = \sum_{i = n}^{N_f} {\langle (\Delta v_{1i})^2 \rangle \over {\lambda_1 - \lambda_i}} \approx  - \sum_{i= n}^{N_f} {{s/N_f} \over {i^2/N_f^2}} \sim -{s N_f \over n}\,.
\ee
Again plugging in the value of $s$, this contribution is equal to our degenerate calculation. This confirms that our splitting of the calculation into degenerate and nondegenerate pieces is sensible. The conclusion is then that for small field displacements, the lowest eigenvalue evolves as 
\be
\langle  \Delta \lambda_1 (s) \rangle \sim - s^{2/3}\,.
\ee

At small $s$, this is a smaller effect than the MMPW estimate, $\delta \lambda \sim - \sqrt{s}$. We would obtain the MMPW result if, instead of dividing the perturbation theory into degenerate and nondegenerate pieces, we had simply done degenerate perturbation theory, allowing all the elements of $\Delta v_{ab}$ to contribute to the new ground state eigenvalue. This is indeed a lower bound on the eigenvalue.

\begin{figure}[t]
	\centering
	\includegraphics[width=\textwidth]{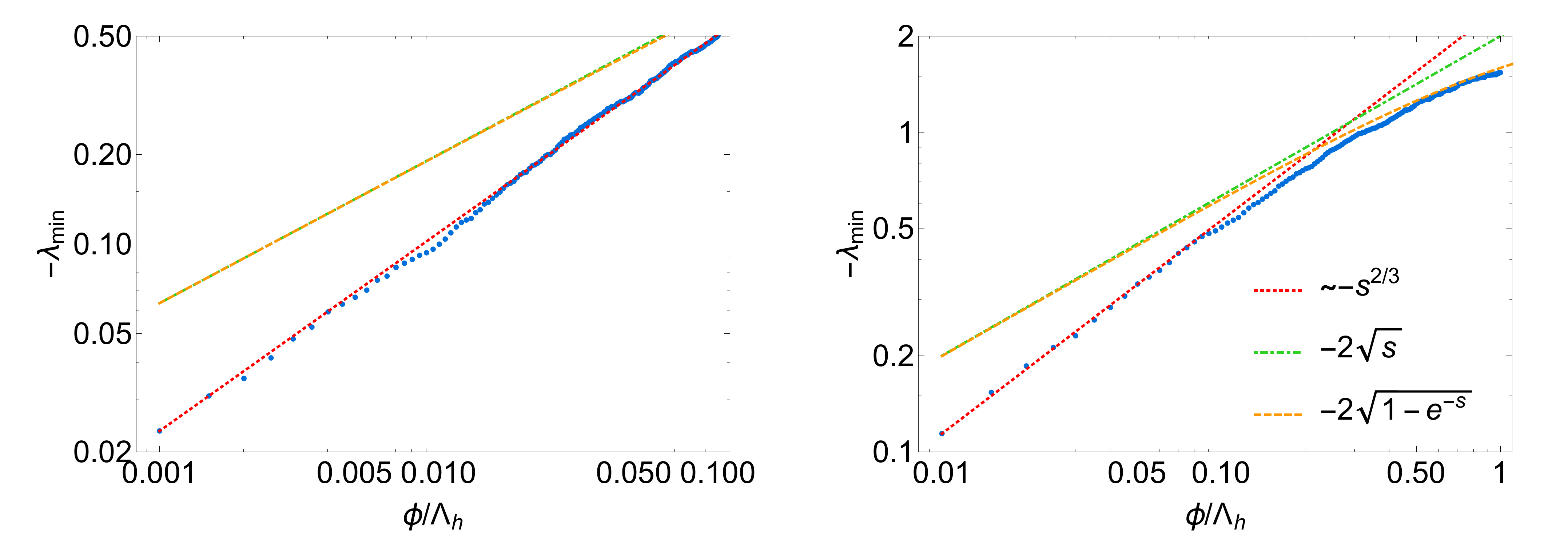}
	\caption{\label{23_vs_12}We plot the approximations $\lambda_{min}\sim-s^{2/3}$ and $\lambda_{min} \sim -2\(1-e^{-s}\)^{1/2} \sim -2s^{1/2}$ compared to numerical realizations of the evolution of the smallest eigenvalue $\lambda_{min}$ of a random matrix with 100 entries. In both cases the matrix has been evolved for 200 steps. The left plot shows the behavior at small field and the good agreement with the $s^{2/3}$ approximation, whereas we see in the right plot that for bigger values of $s$, the approximation of MMPW works better. Since our analytical method does not allow us to compute the coefficient in front of $s^{2/3}$, we normalized it to match the first value of our simulation.}
\end{figure}

In figure \ref{23_vs_12} we show our analytical result together with the results of MMPW and simulations, for both small and large field displacements. The $\lambda_{1}=\lambda_{min}\sim -s^{2/3}$ estimate works decisively better for small $s$, whereas for $s\sim1$ the simulation saturates the bound given by $\lambda\sim -2(1-e^{-s})^{1/2}$. In the rest of the paper we will leave the exponent of $s$ as a parameter in order to study how our results depend on it and we will show results with both estimates of the mass evolution.

 
\subsection{Very small $s$}
When the field displacement is very small, there is no need to do degenerate perturbation theory, and we can do a simpler analysis. The completely nondegenerate regime occurs when the size of the degenerate matrix calculated above, $n \sim N_f s^{1/3}$ is smaller than order 1; in other words, $s \lesssim N_f^{-3}$. In this regime, we can calculate using only nondegenerate perturbation theory, obtaining
\be
\Delta \lambda_1 = \sum_{i=2}^{N_f}  {v_{1i}^2 \over {\lambda_1 - \lambda_i}}~,
\ee
which leads to 
\be
\left< \Delta \lambda_1 \right> \sim  \sum_{i=2}^{N_f}  {s/N_f \over {i^2/N_f^2}} \sim  s N_f \ \ \ {\rm for } \ s \lesssim N_f^{-3}\,.
\ee 

 
\subsection{Variance of the eigenvalue evolution}\label{ssec:} 

To fully characterize the eigenvalue evolution, we want to also calculate the variance of the change in eigenvalue, $\langle (\Delta \lambda_1)^2 \rangle$, to understand whether different realizations will deviate significantly from the average $s^{2/3}$ evolution. We do this separately for the degenerate and nondegenerate contribution. For any realization, the total change in eigenvalue is given by the sum of a degenerate and non-degenerate contribution, 
\be
\Delta \lambda_1 = \Delta \lambda_1^{\rm deg} + \Delta \lambda_1^{\rm non-deg} \,.
\ee
Since the matrix elements appearing in the degenerate contribution are statistically independent of the matrix elements appearing in the non-degenerate contribution, the total variance is the sum of the variance for each contribution. 

For the degenerate contribution, the probability distribution for the smallest eigenvalue is determined by the statistics of the $n\times n$ submatrix. These eigenvalues obey the Wigner semicircle law. The smallest eigenvalue has a variance $n^{-4/3}$ times the square of the expected smallest eigenvalue \cite{dalla}, so the degenerate part contributes

 \be
 {\left<  \left(\Delta \lambda_1^{\rm deg} - \langle \Delta \lambda_1^{\rm deg}\rangle \right)^2   \right>  \over {\left< \Delta \lambda_1 \right>^2} }
 \sim {1 \over n^{4/3} } \,.
\ee

For the non-degenerate part, we can calculate the variance simply by
\be
\left<  \left(\Delta \lambda_1^{\rm non-deg} - \langle \Delta \lambda_1^{\rm non-deg}\rangle \right)^2   \right> =
   \sum_{i = n}^{N_f} {\left< \left[ (\Delta v_{1i})^2 - \langle \Delta v_{1i}\rangle^2 \right]^2 \right> \over \left(\lambda_1 - \lambda_i\right)^2}\,.
\ee
Since $\Delta v_{1i}$ is a Gaussian random variable with variance $s/N_f$, this becomes
\be
\left<  \left(\Delta \lambda_1^{\rm non-deg} - \langle \Delta \lambda_1^{\rm non-deg}\rangle \right)^2   \right> =   \sum_{i = n}^{N_f} {s^2/N_f^2 \over \left(\lambda_1 - \lambda_i\right)^2}~.
\ee
Using again that $\lambda_i - \lambda_1 \sim i^2/N_f^2$ gives
\be
\left<  \left(\Delta \lambda_1^{\rm non-deg} - \langle \Delta \lambda_1^{\rm non-deg}\rangle \right)^2   \right> =   {s^2 N_f^2 \over n^3} \sim  {s \over N_f}\,,
\ee
where in the last simequality we have used $n \sim s^{1/3} N_f$.

The quantity of interest is the ratio between the variance and the square of the mean,
\be
 { \left<  \left(\Delta \lambda_1^{\rm non-deg} - \langle \Delta \lambda_1^{\rm non-deg}\rangle \right)^2   \right>  \over {\left< \Delta \lambda_1 \right>^2} }\sim {1 \over N_f s^{1/3}}\,.
\ee
This ratio is small in the regime where degenerate perturbation theory is important, $s \gtrsim N_f^{-3}$. Therefore, in the degenerate regime, $s \gtrsim N_f^{-3}$, the spread in the eigenvalue distribution is small compared to its mean. In the completely non-degenerate regime, $s \lesssim N_f^{-3}$, the variance is of the same order as the mean or larger.

Normally, this would justify replacing the eigenvalue with its mean value, as we have done in the following analysis. However, it may be the case that conditioning on a sufficiently long period of inflation selects  the tails of the distribution, where the second derivative increases more slowly.\footnote{We thank David Marsh for pointing this out and for discussions on the importance of the variance.} We leave this complication for future work.


\section{Predictions for the spectral tilt} \label{tilt_comp}

In this section we focus on the inflation near a critical point, as a natural characterization of a small initial gradient. It is in principle possible that the gradient is small and yet there is no critical point around the point of interest. We comment on this possibility in the next subsection.
Near a critical point, the problem then reduces to calculating the perturbations in single field hilltop inflation with a modified mass. We will treat the case of a general monomial dependence on the field, i.e.~the potential
\be \label{mod_hilltop}
	V = V_0 - \(\frac{m_0^2}{2}+c\phi^p\)\phi^2  \, ,
\ee
where now $c = \sqrt{N_f}\Lambda_v^4/\Lambda_h^{p+2} \times \mathcal{O}(1)$. The case in equation \eqref{hilltop_potential} is recovered for $p=1/2$, but we want to be able to treat the $p=2/3$ case and study the dependence on $p$, looking towards possible generalizations of the model. We are particularly interested in computing the tilt of the power spectrum $n_s$, or better to find a generic upper bound on it.

In order to do so we start computing the slow-roll parameters. In hilltop models, inflation happens near the top of the hill, so that usually $\eta_V\gg\epsilon_V$. In our case
\be
	\eta_V\simeq-\frac{1}{3H^2}\(m_0^2 + c(p+2)(p+1)\phi^p\) \, .
\ee
From this the tilt of the power spectrum follows
\be
	n_s-1 = 2 \left.\eta_V\right|_* - 6 \left.\epsilon_V\right|_* \lesssim -\frac{2}{3H^2}\(m_0^2+c(p+2)(p+1)\phi^p_*\) \, ,
\ee
where $|_*$ means that the quantity needs to be evaluated as the relevant mode exits the Hubble radius. We can determine the value of the field in terms of the number of efolds inflation lasted
\be
	N_e &= & \int_{\phi_i}^{\phi_f}\textrm{d}\phi\frac{H}{\dot{\phi}} \simeq -3H^2\int_{\phi_i}^{\phi_f}\frac{\textrm{d}\phi}{V'} \\
	& = & \frac{3H^2}{p m_0^2}\[\ln\frac{\phi^p}{m_0^2+c(p+2)\phi^p}\]_{\phi_i}^{\phi_f} \, .
\ee

The $\phi^2$ regime dominates the system at small field values, whereas the $\phi^{p+2}$ regime takes over for
\be
	\phi^p \gtrsim \frac{m_0^2}{2c} \, .
\ee
Since the mass term must be small to allow for inflation, while the coefficient $c$ is not small, inflation ends well in the $\phi^{p+2}$ regime. So we approximate
\be \label{approx}
	\phi_f^p \gg \frac{m_0^2}{2c} \gtrsim \frac{m_0^2}{(2+p)c} \, .
\ee
We check in appendix \ref{A} that this approximation is self-consistent, and indeed inflation ends well in the $\phi^p$ regime.

This gives the number of efoldings
\be \label{N_efolds}
	N_e = \frac{3H^2}{p m_0^2} \ln\[\frac{m_0^2+c(p+2)\phi_*^p}{c(p+2)\phi_*^p}\] \, ,
\ee
where we replaced $\phi_i$ by $\phi_*$. We can now invert this result and find the tilt of the power spectrum to be
\be \label{bound}
	n_s-1 \lesssim -\frac{2\mu^2}{3}\frac{\exp(\mu^2 p N_e/3)+p}{\exp(\mu^2 p N_e/3)-1} \, ,
\ee
in terms of $\mu\equiv m_0/H$.

\begin{figure}[t]
	\centering
	\includegraphics[width=.9\textwidth]{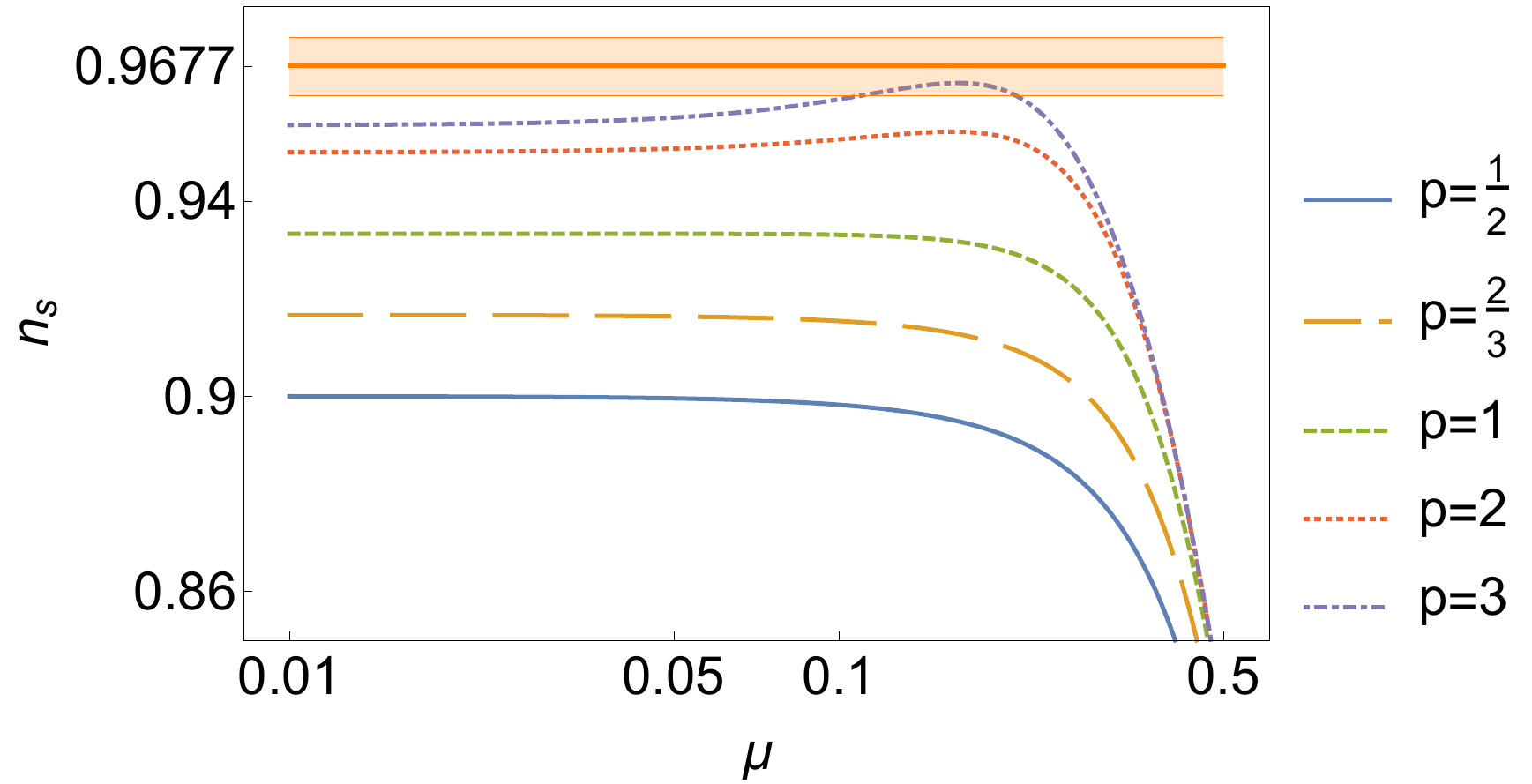}
	\caption{\label{ns_vs_mu}We show the dependence of the scalar spectrum $n_s$ on the ratio $\mu = m_0/H$ for 60 efolds and different values of $p$ in equation \eqref{mod_hilltop}, together with the measurement from Planck \cite{Ade:2015lrj} at 68\% confidence level.}
\end{figure}

This is one of the main results of this paper. One can easily estimate that for either $p=1/2$ or $p=2/3$ and any ratio of initial mass and Hubble constant, at 60 efolds the spectral tilt predicted by this model is excluded by observation. In figure \ref{ns_vs_mu} we plotted the $\mu$ dependence of $n_s$ for different values of $p$ together with the best value from Planck \cite{Ade:2015lrj}.

The reader might be puzzled about how to recover the quadratic hilltop model from our treatment, as one would imagine to recover it in the limit $p\to0$. In fact, that is the limit in which the mass changes instantaneously to some big negative value, thus preventing inflation in the first place. On the other hand, the limit $p\gg1$ is the correct one to recover the hilltop quadratic model: for small field values, large $p$ suppresses the second term in the mass squared in equation \eqref{mod_hilltop}. We can see in figure \ref{ns_vs_mu} that indeed the model agrees with data for large values of $p$.

In light of this, one might still hope to generalize the MMPW model by embedding different statistics in order to obtain a larger $p$ in the effective approach and find agreement with data. Bear in mind that a higher $p$ corresponds to a \textit{smaller} change in the mass. It is hard (perhaps impossible) to build a random matrix model for which eigenvalue repulsion is small when the eigenvalues are close and it increases as they move apart. This would be the case for any $p>1$ and we would then need to introduce some new restoring force as well, in order to avoid infinite drifting of the eigenvalues.

We expect that a small modification of the evolution of the Hessian would not alter our result significantly. Maintaining the same approach as in MMPW, one would still be able to find an approximate single field description with a modified mass, but perhaps with a different dependence on $\phi$.

 
\subsection{Away from the hilltop}\label{ssec:4p1}

It is possible to have a slow-roll flat region of the potential that is \textit{not} near a critical point. In this case, the model in \eqref{mod_hilltop} needs to be generalized to include a gradient
\be \label{pot_lin}
	V = V_0 + \gamma \phi - \(\frac{m_0^2}{2}+c\phi^p\)\phi^2  \, ,
\ee
where $ \gamma$ is a constant. Analytical solutions are harder to find with this extra term. We numerically tested a few values of nonzero $\gamma$ and found that some values can indeed recover an acceptable value of $n_s$. For completeness we report one set of parameters that lead to a tilt in agreement with observations in appendix \ref{B}.

Unfortunately, as we will explain in section \ref{ssec-gradient}, the probability distribution of the gradient is ill-defined in an MMPW landscape. If such global statistics existed, we could have quantified the probability of having acceptable values of $\gamma$. More generally, we could have estimated the number of regions with a small gradient but not near a critical point, thus not included in our near-critical-point analysis. In a globally defined potential that is not constructed by the stochastic MMPW mapping, those regions are rare in an appropriate choice of variables. Thus we think our analysis near critical points still has its merit. We will leave it to future work to consolidate global statistics with the MMPW model.


\section{Unsatisfactory aspects of the model}\label{issues}

In this section, we point out two unsatisfactory aspects of the MMPW model, namely the incorrect modeling of very small field displacements and the growth without bound of the variance of gradient.


\subsection{Problems with very small field displacements}

The basic assumption of MMPW is that for small field displacement, the matrix of second derivatives evolves statistically, with variance
\begin{equation}
\langle (\delta v_{ab})^2 \rangle = {1 \over N_f} (1 + \delta_{ab} ) s\,.
\end{equation}
This means that the size of the elements of the matrix depends on $s$ as $\delta v_{ab} \sim \sqrt{s}$. 

However, in a potential with nonsingular third derivatives, the evolution of the second derivatives is governed by the third derivative matrix,
\be
\delta v_{ab} = v_{abc} \frac{\delta \phi^c}{\Lambda_{h}}\,.
\ee
Moving in any particular direction, this gives
\be
\delta v_{ab} \sim s \,.
\ee
 In other words, the elements of the Hessian must change linearly with the field displacement, for sufficiently small field displacement. We would expect this linear behavior to be valid until either the field changes direction, so that different elements of $v_{abc}$ become relevant, or until the field moves far enough that the matrix of third derivatives changes. On these longer length scales, it is reasonable to model the evolution of the Hessian as a diffusion process, as in MMPW.  However, on shorter scales, the diffusion behavior $\delta v_{ab} \sim \sqrt{s}$ gives {\it too large} a change in the second derivative. Since the changing second derivative is the essence of the conflict with the observed spectral index, it would be interesting to try to improve the model in this regard. 



\subsection{Problems with global statistics on vacuum energy and gradient}
\label{ssec-gradient}

In section \ref{ssec:4p1}, we have explained that the predictions also depends on the value of the gradient $\gamma$, which was ignored in the near-hilltop analysis. Since the model assumes that the potential is statistically homogeneous in field space, the natural procedure is to find a probability distribution for these quantities that is left invariant under evolution in field space, as was done for the Hessian. 

Unfortunately, as we will now explain, there is no normalizable probability distribution for the gradient that is invariant under the MMPW evolution equations. This issue has been raised previously in  \cite{MarshPajer, Battefeld16}. 
 In the MMPW model, the gradient evolves according to
\be
\delta v_a = v_{ab} {\delta \phi^b \over \Lambda_h}\,.
\ee
One can already see the problem: as described in equation (\ref{evolution}), the statistics of the second derivative are independent of the value of the gradient. Therefore, $v_a$ is equally likely to increase or decrease and its variance grows without a bound.

We can attach equations to this argument as follows. Although for the purposes of inflation one chooses a path through field space determined by gradient descent, we are free to follow any path in order to analyze properties of the potential. For simplicity, we choose to evolve the gradient as we move along the $\phi^1$ direction, which we parameterize as $ s=\phi_{1}/\Lambda $. Focusing on $v_1$ for simplicity, we can integrate the evolution equation to obtain
\be
v_1(s) = v_1(0) + \int_0^{s} ds' \  v_{11}(s')
\ee
In this equation, all quantities depend also on the other fields $\phi_i$'s, but since we are only moving in the $\phi_1$ direction, we suppress the other fields.

Now we can calculate 
\be
\left\langle \left( v_1(s) - v_1(0) \right)^2 \right\rangle 
&=& 
\int_0^{s} ds' \int_0^{s}ds'' \ \left\langle v_{11}(s') \ v_{11}(s'') \right\rangle
\\ \nonumber
&=& 2 \int_0^{s} ds_+ \int_{-s_{+}}^{s_{+}} ds_- 
\left\langle v_{11}(s_++s_{-}) v_{11}(s_+-s_{-}) \right\rangle
\ee
For large $s$, the memory of the initial conditions is lost, and the 2-point function depends only on the separation $\left( s' - s'' \right)/2=s_-$. Therefore, the large $s$ behavior is captured by
\be
\left\langle \left( v_1(s) - v_1(0) \right)^2 \right\rangle \approx 2  \int_0^{s} ds_+ \int_{-s_+}^{s_+} ds_- 
G(s_{-}) \,,
\ee
 The correlator $G(s_-)$ is positive and its integral is finite, so the integral grows linearly without bound as $s\propto \phi_1$ increases, 
 \be
 \left\langle \left( v_1(s) - v_1(0) \right)^2 \right\rangle  \sim s
 \ee
As a result, no probability distribution with a finite variance can be left invariant by the evolution.

One can make a similar argument for the height of the potential $v_0$; however, the two-point function of the gradient will now appear on the right side of the equation. Since this is itself undefined, one is not in a position to determine what probability distribution, if any, is left invariant under the evolution.

\section{Discussion} \label{discussion}

We employed the MMPW model \cite{Marsh:2013qca} for the landscape to study inflation in the limit of a large number of fields. We showed that inflation near a critical point can happen, but it generically leads to a scalar spectral index that is too red and therefore in conflict with current observations. We stress that the universes predicted by this model seem to allow intelligent life to develop, since a slightly redder tilted power spectrum does not threaten our existence. Therefore, the discrepancy with data cannot be argued away using anthropic arguments.

Dropping the assumption that a critical point is nearby, therefore allowing for a gradient term, we found cases that are consistent with current data. The sets of parameters that work are few and far between, adopting the simplest statistics for the gradient. Unfortunately, the MMPW model does not provide a prescription to determine the probability distribution for the gradient of the potential. Hence, while we suspect these sets are rare, we could not quantify their typicality. 

The apparent incompatibility with observations of the MMPW model leads us to three possible options: we need better statistical model for a random landscape; our universe is not typical; or the kinematics of a random multifield landscape alone does not describe our universe.

We believe that the first option is disfavored. As described in section \ref{tilt_comp}, eigenvalue repulsion leads to a steepening of the potential, resulting in a spectral index in conflict with observations. The conflict persists if the eigenvalue repulsion is artificially weakened or strengthen, see figure \ref{ns_vs_mu}. Therefore we expect that modifications of the dynamical evolution of the Hessian, which generalize the MMPW model, would still fail to agree with data. 

The second option deprives us from a powerful weapon, namely naturalness. In this case, statistical tools are not useful in describing the universe. We want to stress that here we do not refer to the fact that slow-roll inflation is difficult to realize in the landscape, as we already knew this to be the case in a random landscape. Rather, our universe is atypical among the ones that undergo inflation and can host sentient life. Giving up the assumption of typicality would prevent $any$ comparison between theory and cosmological observations.

We speculate that the third option is therefore the most promising. We interpret it as pointing toward the fact that extra structures in the landscape have played an important role in our cosmological history. For example, some fields enjoy symmetries (e.g.~a shift symmetry) that make the potential in those flatter than in the fully random case. Even if these shift-symmetric field only form a small subset of all the fields, a theory of initial conditions\footnote{A concrete example is that a tunnelling initial condition favors unwinding inflation \cite{D'Amico:2012sz,D'Amico:2012ji}, since the end point of a tunnelling event is a natural starting point of such inflation path.} might favor inflationary paths in those directions. This option is supported by other hints in the literature: due to eigenvalue repulsion, critical points in a random multifield landscape are very unlikely to be minima \cite{wasteland, gary, douglas, ana}; if a minimum does exist, it is likely to have a rapid nonperturbative decay rate \cite{greeneinst, dine2, dine}. Taken together with our results, one might reasonably conclude that describing our universe by a theory of many fields without any structure is doomed to failure.

\section*{Acknowledgments}

It is a pleasure to thank David Marsh, Liam McAllister, Francisco Pedro and Alex Westphal for interesting discussions. E.P.~and R.G.~are supported by the D-ITP consortium, a program of the Netherlands Organization for Scientific Research (NWO) that is funded by the Dutch Ministry of Education, Culture and Science~(OCW).


\appendix

\section{Approximations used in section \ref{tilt_comp}}\label{A}

We discuss here the validity of the assumptions we made in section \ref{tilt_comp}. First we will show that inflation does indeed end in the $\phi^p$ regime as postulated in our calculation. We will then justify the assumptions $H=\textrm{const.}$ and $\phi_f^p\gg m_0^2/c$, \emph{i.e}.~that $V_0$ dominates and inflation ends \emph{well} in the $\phi^p$ regime.

\subsection{Inflation ends in the $\phi^p$ regime} \label{A1}

Inflation ends when the slow-roll parameter $\epsilon\sim1$ and for the model in \eqref{mod_hilltop}, this can be approximated by
\be \label{epsilon}
	\epsilon_V = \frac{M_p^2}{2}\(\frac{m_0^2\phi + c(p+2)\phi^{p+1}}{V_0}\)^2 \, .
\ee

Within our approximation, the condition for ending inflation is then
\be \label{end_inf}
	s_f^{p+1}\frac{M_p}{\Lambda_h}\frac{\sqrt{N_f}(p+2)}{v_0\sqrt{N_f}} \simeq 1 \, ,
\ee
where we used that $c \sim \sqrt{N_f}\Lambda_v^4/\Lambda_h^{p+2} $ and defined $s$ as in \eqref{rescaled}. Morever, we should keep in mind that with our conventions $v_0\sqrt{N_f}\sim1$, so the end of inflation occurs at 
\be
s_f^{p+1} \sim {\Lambda_h \over M_P} {1 \over \sqrt N_f}\,.
\ee
Since we assume $\Lambda_h < M_P$, inflation always ends at small $s$, so understanding the field evolution for small field displacement $s$ is sufficient. 

To check that the $\phi^p$ term dominates at the end of inflation, we compare now the first and second terms of the numerator in \eqref{epsilon}. The ratio of the quadratic term to the $\phi^p$ term can be simplified, by a quite tedious and not enlightening series of manipulations to
\be \label{ratio}
	\frac{m_0\phi_f}{c(p+2)\phi_f^{p+1}} = \mu^2\(\frac{\Lambda_h}{M_p}\)^{\frac{p+2}{p+1}}\(\frac{v_0\sqrt{N_f}}{(p+2)\sqrt{N_f}}\)^{\frac{2p+1}{p+1}} \, .
\ee
The last term in \eqref{ratio} is always smaller than 1 (barring unusual fluctuations of $v_0$)
The conditions
\be
	\begin{split}
	\mu & < 1 \,,\\
	\frac{\Lambda_h}{M_p} & < 1\,,
	\end{split}
\ee
are sufficient to justify our approximation and are consistent with our initial assumptions.

\subsection{No need for more stringent assumptions}

In section \ref{tilt_comp} we assumed $H=\textrm{const.}$ and in section \ref{A1} we showed $\phi_f^p\gg m_0^2/c$ to hold true in our model. In general, these conditions are not enough to ensure that our results are correct; as an example consider the choice of parameters
\be
	p = 2/3 \qquad V_0 = 10^{-6}M_P^4 \qquad m_0 = 10^{-6}M_P \qquad c = 10^{-8}M_P^{2-p} \, .
\ee
They do not violate any of the assumptions we made, but still they violate the bound in equation \eqref{bound}. The second assumption is indeed not quite enough to ensure the validity of our results, but rather we would need to assume the more stringent $\phi_f\gg\phi_*$. However we will show here that for our parametrization and typical values for the parameters, this requirement is already automatically satisfied. With non-atypical values we mean
\be
	V_0\simeq\Lambda_v^4 \qquad c\simeq\sqrt{N_f}\frac{\Lambda_v^4}{\Lambda_h^{2+p}} \, .
\ee



First of all we need to find $\phi_f$. Inflation ends when the slow roll parameters become $\sim1$ and in our model $\eta$ becomes $\sim1$ before $\epsilon$, therefore $\eta\sim1$ is the stronger condition under which to prove $\phi_f/\phi_*\gg1$.

Assuming inflation ends in the $\phi^{p+2}$ regime and imposing $\eta=1$, one gets
\be
	\phi_f^p = \frac{3H^2}{c(p+2)(p+1)} \, .
\ee

Let's assume that $\phi_f\gg\phi_*$ and check for consistency of equation \eqref{N_efolds}. I invert equation \eqref{N_efolds} and take the limit $m_0\to0$ and find\footnote{Notice that $\phi_*$ would be even smaller for $m_0\neq0$ as it would be exponentially suppressed}
\be
	\phi_*^p = \frac{3H^2}{c(p+2)pN_e} \, .
\ee

It is now straightforward to take the ratio:
\be
	\frac{\phi_*}{\phi_f} = \(\frac{p+1}{pN_e}\)^{1/p} = \(\frac{1}{24}\)^{3/2}\simeq 8\times10^{-3}\ll1 \, ,
\ee
showing that we do not need stronger assumptions in our calculations.\\

\section{A working set of values introducing $\gamma$} \label{B}

We mentioned in section \ref{ssec:4p1} that adding a linear term to the potential which is sizable at the beginning of inflation, as in equation \eqref{pot_lin}, leads to a universe compatible with observations for some particular values of the parameters. For completeness, we report in this appendix one viable set of parameters, which we found through numerical investigation.

We assume the number of fields to be $N_f=100$ and we give values to the vertical and horizontal scales of our potential: $\Lambda_v=0.3\times10^{-4}M_P$ and $\Lambda_h=M_P$. Then the set of parameters
\be
\begin{split}
	V_0 = \Lambda_v^4\,,\quad \gamma = 1.3\times 10^3 \frac{\Lambda_v^6}{M_P^3}\,,\quad m_0 = 10^{-3/2} N_f^{1/4}\frac{\Lambda_v^2}{\Lambda_h} \,,\quad c = N_f^{1/2}\frac{\Lambda_v^4}{\Lambda_h^{2+p}} \, ,
\end{split}
\ee
with $p=2/3$, produces a power spectrum roughly in agreement with data. Inflation lasts $N\gtrsim60$ and the power spectrum intensity and tilt is
\be
\begin{split}
	\Delta^2 &\simeq 4.7 \times 10^{-9} \quad \text{and}\quad	n_s &\simeq 0.96 \, .
\end{split}
\ee

As one can see, the amplitude of the power spectrum is close to the observed one (up to a factor of 2) and its tilt is in agreement with data, thus defying the problems encountered in section \ref{tilt_comp}.

\bibliographystyle{utphys}
\bibliography{random}

\providecommand{\href}[2]{#2}\begingroup\raggedright\begin{thebibliography}{10}

\bibitem{Guth:1980zm}
A.~H. Guth, ``{The Inflationary Universe: A Possible Solution to the Horizon
  and Flatness Problems},''
\href{http://dx.doi.org/10.1103/PhysRevD.23.347}{{\em Phys. Rev.} {\bfseries
  D23} (1981) 347--356}.

\bibitem{Linde:1981mu}
A.~D. Linde, ``{A New Inflationary Universe Scenario: A Possible Solution of
  the Horizon, Flatness, Homogeneity, Isotropy and Primordial Monopole
  Problems},''
\href{http://dx.doi.org/10.1016/0370-2693(82)91219-9}{{\em Phys. Lett.}
  {\bfseries B108} (1982) 389--393}.

\bibitem{Albrecht:1982wi}
A.~Albrecht and P.~J. Steinhardt, ``{Cosmology for Grand Unified Theories with
  Radiatively Induced Symmetry Breaking},''
\href{http://dx.doi.org/10.1103/PhysRevLett.48.1220}{{\em Phys. Rev. Lett.}
  {\bfseries 48} (1982) 1220--1223}.

\bibitem{Ade:2015lrj}
{\bfseries Planck} Collaboration, P.~A.~R. Ade {\em et~al.}, ``{Planck 2015
  results. XX. Constraints on inflation},''
\href{http://arxiv.org/abs/1502.02114}{{\ttfamily arXiv:1502.02114
  [astro-ph.CO]}}.

\bibitem{BP}
R.~Bousso and J.~Polchinski, ``Quantization of four-form fluxes and dynamical
  neutralization of the cosmological constant,'' {\em JHEP} {\bfseries 06}
  (2000) 006,
\href{http://arxiv.org/abs/hep-th/0004134}{{\ttfamily hep-th/0004134}}.

\bibitem{Wigner:rm}
E.~P. Wigner, ``{On the statistical distribution of the widths and spacings of
  nuclear resonance levels},''
{\em Mathematical Proceedings of the Cambridge Philosophical Society}
  {\bfseries 47} (1951) no. 04 790--798.

\bibitem{Dyson:bm}
F.~J. Dyson, ``{A Brownian-Motion model for the Eigenvalues of a Random
  Matrix},''
{\em J. Math. Phys.} {\bfseries 3} (1962) 1191.

\bibitem{Easther:2005zr}
R.~Easther and L.~McAllister, ``{Random matrices and the spectrum of
  N-flation},'' \href{http://dx.doi.org/10.1088/1475-7516/2006/05/018}{{\em
  JCAP} {\bfseries 0605} (2006) 018},
\href{http://arxiv.org/abs/hep-th/0512102}{{\ttfamily arXiv:hep-th/0512102
  [hep-th]}}.

\bibitem{Denef:2004ze}
F.~Denef and M.~R. Douglas, ``{Distributions of flux vacua},''
  \href{http://dx.doi.org/10.1088/1126-6708/2004/05/072}{{\em JHEP} {\bfseries
  05} (2004) 072},
\href{http://arxiv.org/abs/hep-th/0404116}{{\ttfamily arXiv:hep-th/0404116
  [hep-th]}}.

\bibitem{MarMcA11}
D.~Marsh, L.~McAllister, and T.~Wrase, ``{The Wasteland of Random
  Supergravities},'' \href{http://dx.doi.org/10.1007/JHEP03(2012)102}{{\em
  JHEP} {\bfseries 1203} (2012) 102},
\href{http://arxiv.org/abs/1112.3034}{{\ttfamily arXiv:1112.3034 [hep-th]}}.

\bibitem{Chen:2011ac}
X.~Chen, G.~Shiu, Y.~Sumitomo, and S.~H.~H. Tye, ``{A Global View on The Search
  for de-Sitter Vacua in (type IIA) String Theory},''
  \href{http://dx.doi.org/10.1007/JHEP04(2012)026}{{\em JHEP} {\bfseries 04}
  (2012) 026},
\href{http://arxiv.org/abs/1112.3338}{{\ttfamily arXiv:1112.3338 [hep-th]}}.

\bibitem{Bachlechner:2012at}
T.~C. Bachlechner, D.~Marsh, L.~McAllister, and T.~Wrase, ``{Supersymmetric
  Vacua in Random Supergravity},''
  \href{http://dx.doi.org/10.1007/JHEP01(2013)136}{{\em JHEP} {\bfseries 01}
  (2013) 136},
\href{http://arxiv.org/abs/1207.2763}{{\ttfamily arXiv:1207.2763 [hep-th]}}.

\bibitem{GreKag13}
B.~Greene, D.~Kagan, A.~Masoumi, D.~Mehta, E.~J. Weinberg, and X.~Xiao,
  ``{Tumbling through a landscape: Evidence of instabilities in
  high-dimensional moduli spaces},''
  \href{http://dx.doi.org/10.1103/PhysRevD.88.026005}{{\em Phys. Rev.}
  {\bfseries D88} no.~2, (2013) 026005},
\href{http://arxiv.org/abs/1303.4428}{{\ttfamily arXiv:1303.4428 [hep-th]}}.

\bibitem{Bachlechner:2014rqa}
T.~C. Bachlechner, ``{On Gaussian Random Supergravity},''
  \href{http://dx.doi.org/10.1007/JHEP04(2014)054}{{\em JHEP} {\bfseries 04}
  (2014) 054},
\href{http://arxiv.org/abs/1401.6187}{{\ttfamily arXiv:1401.6187 [hep-th]}}.

\bibitem{McARen12}
L.~McAllister, S.~Renaux-Petel, and G.~Xu, ``{A Statistical Approach to
  Multifield Inflation: Many-field Perturbations Beyond Slow Roll},''
\href{http://arxiv.org/abs/1207.0317}{{\ttfamily arXiv:1207.0317
  [astro-ph.CO]}}.

\bibitem{Yan12a}
I.-S. Yang, ``{Probability of Slowroll Inflation in the Multiverse},''
  \href{http://dx.doi.org/10.1103/PhysRevD.86.103537}{{\em Phys.Rev.}
  {\bfseries D86} (2012) 103537},
\href{http://arxiv.org/abs/1208.3821}{{\ttfamily arXiv:1208.3821 [hep-th]}}.

\bibitem{Marsh:2013qca}
M.~C.~D. Marsh, L.~McAllister, E.~Pajer, and T.~Wrase, ``{Charting an
  Inflationary Landscape with Random Matrix Theory},''
  \href{http://dx.doi.org/10.1088/1475-7516/2013/11/040}{{\em JCAP} {\bfseries
  1311} (2013) 040},
\href{http://arxiv.org/abs/1307.3559}{{\ttfamily arXiv:1307.3559 [hep-th]}}.

\bibitem{Pedro:2013nda}
F.~G. Pedro and A.~Westphal, ``{The Scale of Inflation in the Landscape},''
  \href{http://dx.doi.org/10.1016/j.physletb.2014.10.022}{{\em Phys. Lett.}
  {\bfseries B739} (2014) 439--444},
\href{http://arxiv.org/abs/1303.3224}{{\ttfamily arXiv:1303.3224 [hep-th]}}.

\bibitem{Dias:2016slx}
M.~Dias, J.~Frazer, and M.~C.~D. Marsh, ``{Simple emergent power spectra from
  complex inflationary physics},''
\href{http://arxiv.org/abs/1604.05970}{{\ttfamily arXiv:1604.05970
  [astro-ph.CO]}}.

\bibitem{Battefeld14}
T.~Battefeld and C.~Modi, ``{Local random potentials of high differentiability
  to model the Landscape},''
  \href{http://dx.doi.org/10.1088/1475-7516/2015/03/010}{{\em JCAP} {\bfseries
  1503} no.~03, (2015) 010},
\href{http://arxiv.org/abs/1409.5135}{{\ttfamily arXiv:1409.5135 [hep-th]}}.

\bibitem{MarshPajer}
E.~Pajer and M.~C.~D. Marsh,
``Unpublished,''.

\bibitem{Battefeld16}
G.~Wang and T.~Battefeld, ``{Random Functions via Dyson Brownian Motion:
  Progress and Problems},''
\href{http://arxiv.org/abs/1607.02514}{{\ttfamily arXiv:1607.02514 [hep-th]}}.

\bibitem{Pedro}
F.~G. Pedro and A.~Westphal, ``{Non-Equilibrium Random Matrix Theory :
  Transition Probabilities},''
\href{http://arxiv.org/abs/1606.07768}{{\ttfamily arXiv:1606.07768
  [cond-mat.stat-mech]}}.

\bibitem{Achucarro:2010jv}
A.~Achucarro, J.-O. Gong, S.~Hardeman, G.~A. Palma, and S.~P. Patil, ``{Mass
  hierarchies and non-decoupling in multi-scalar field dynamics},''
  \href{http://dx.doi.org/10.1103/PhysRevD.84.043502}{{\em Phys. Rev.}
  {\bfseries D84} (2011) 043502},
\href{http://arxiv.org/abs/1005.3848}{{\ttfamily arXiv:1005.3848 [hep-th]}}.

\bibitem{Yang:2012bs}
I.-S. Yang, ``{The Strong Multifield Slowroll Condition and Spiral
  Inflation},'' \href{http://dx.doi.org/10.1103/PhysRevD.85.123532}{{\em Phys.
  Rev.} {\bfseries D85} (2012) 123532},
\href{http://arxiv.org/abs/1202.3388}{{\ttfamily arXiv:1202.3388 [hep-th]}}.

\bibitem{Ahlqvist:2013wla}
P.~Ahlqvist, B.~Greene, and D.~Kagan, ``{Exploring Spiral Inflation in String
  Theory},''
\href{http://arxiv.org/abs/1308.0538}{{\ttfamily arXiv:1308.0538 [hep-th]}}.

\bibitem{Achucarro:2010da}
A.~Achucarro, J.-O. Gong, S.~Hardeman, G.~A. Palma, and S.~P. Patil,
  ``{Features of heavy physics in the CMB power spectrum},''
  \href{http://dx.doi.org/10.1088/1475-7516/2011/01/030}{{\em JCAP} {\bfseries
  1101} (2011) 030},
\href{http://arxiv.org/abs/1010.3693}{{\ttfamily arXiv:1010.3693 [hep-ph]}}.

\bibitem{Achucarro:2012sm}
A.~Achucarro, J.-O. Gong, S.~Hardeman, G.~A. Palma, and S.~P. Patil,
  ``{Effective theories of single field inflation when heavy fields matter},''
  \href{http://dx.doi.org/10.1007/JHEP05(2012)066}{{\em JHEP} {\bfseries 05}
  (2012) 066},
\href{http://arxiv.org/abs/1201.6342}{{\ttfamily arXiv:1201.6342 [hep-th]}}.

\bibitem{Avgoustidis:2012yc}
A.~Avgoustidis, S.~Cremonini, A.-C. Davis, R.~H. Ribeiro, K.~Turzynski, and
  S.~Watson, ``{Decoupling Survives Inflation: A Critical Look at Effective
  Field Theory Violations During Inflation},''
  \href{http://dx.doi.org/10.1088/1475-7516/2012/06/025}{{\em JCAP} {\bfseries
  1206} (2012) 025},
\href{http://arxiv.org/abs/1203.0016}{{\ttfamily arXiv:1203.0016 [hep-th]}}.

\bibitem{mehta}
M.~L. Mehta, ``{Random Matrices},'' {\em No.~142 in Pure and Applied
  Mathematics} {\bfseries Academic Press} (204) .

\bibitem{dalla}
S.~{Dallaporta}, ``{Eigenvalue variance bounds for Wigner and covariance random
  matrices},'' {\em ArXiv e-prints} (Mar., 2012) ,
  \href{http://arxiv.org/abs/1203.1597}{{\ttfamily arXiv:1203.1597 [math.PR]}}.

\bibitem{D'Amico:2012sz}
G.~D'Amico, R.~Gobbetti, M.~Schillo, and M.~Kleban, ``{Inflation from Flux
  Cascades},'' \href{http://dx.doi.org/10.1016/j.physletb.2013.07.050}{{\em
  Phys. Lett.} {\bfseries B725} (2013) 218--222},
\href{http://arxiv.org/abs/1211.3416}{{\ttfamily arXiv:1211.3416 [hep-th]}}.

\bibitem{D'Amico:2012ji}
G.~D'Amico, R.~Gobbetti, M.~Kleban, and M.~Schillo, ``{Unwinding Inflation},''
  \href{http://dx.doi.org/10.1088/1475-7516/2013/03/004}{{\em JCAP} {\bfseries
  1303} (2013) 004},
\href{http://arxiv.org/abs/1211.4589}{{\ttfamily arXiv:1211.4589 [hep-th]}}.

\bibitem{wasteland}
D.~Marsh, L.~McAllister, and T.~Wrase, ``{The Wasteland of Random
  Supergravities},'' \href{http://dx.doi.org/10.1007/JHEP03(2012)102}{{\em
  JHEP} {\bfseries 03} (2012) 102},
\href{http://arxiv.org/abs/1112.3034}{{\ttfamily arXiv:1112.3034 [hep-th]}}.

\bibitem{gary}
X.~Chen, G.~Shiu, Y.~Sumitomo, and S.~H.~H. Tye, ``{A Global View on The Search
  for de-Sitter Vacua in (type IIA) String Theory},''
  \href{http://dx.doi.org/10.1007/JHEP04(2012)026}{{\em JHEP} {\bfseries 04}
  (2012) 026},
\href{http://arxiv.org/abs/1112.3338}{{\ttfamily arXiv:1112.3338 [hep-th]}}.

\bibitem{douglas}
M.~R. Douglas, \href{http://dx.doi.org/10.1142/9789814412551_0012}{``{The
  String landscape and low energy supersymmetry},''} in {\em Strings, gauge
  fields, and the geometry behind: The legacy of Maximilian Kreuzer},
  A.~Rebhan, L.~Katzarkov, J.~Knapp, R.~Rashkov, and E.~Scheidegger, eds.,
  pp.~261--288.
\newblock 2012.
\newblock
\href{http://arxiv.org/abs/1204.6626}{{\ttfamily arXiv:1204.6626 [hep-th]}}.
\newblock

\bibitem{ana}
A.~Achúcarro, P.~Ortiz, and K.~Sousa, ``{A new class of de Sitter vacua in
  String Theory Compactifications},''
\href{http://arxiv.org/abs/1510.01273}{{\ttfamily arXiv:1510.01273 [hep-th]}}.

\bibitem{greeneinst}
B.~Greene, D.~Kagan, A.~Masoumi, D.~Mehta, E.~J. Weinberg, and X.~Xiao,
  ``{Tumbling through a landscape: Evidence of instabilities in
  high-dimensional moduli spaces},''
  \href{http://dx.doi.org/10.1103/PhysRevD.88.026005}{{\em Phys. Rev.}
  {\bfseries D88} no.~2, (2013) 026005},
\href{http://arxiv.org/abs/1303.4428}{{\ttfamily arXiv:1303.4428 [hep-th]}}.

\bibitem{dine2}
M.~Dine and S.~Paban, ``{Tunneling in Theories with Many Fields},''
  \href{http://dx.doi.org/10.1007/JHEP10(2015)088}{{\em JHEP} {\bfseries 10}
  (2015) 088},
\href{http://arxiv.org/abs/1506.06428}{{\ttfamily arXiv:1506.06428 [hep-th]}}.

\bibitem{dine}
M.~Dine, ``{Classical and Quantum Stability in Putative Landscapes},''
\href{http://arxiv.org/abs/1512.08125}{{\ttfamily arXiv:1512.08125 [hep-th]}}.

\end{thebibliography}\endgroup

\end{document}